\documentclass[sigconf,nonacm]{acmart}
\usepackage{multirow}
\usepackage{booktabs}
\usepackage{xcolor}
\usepackage{colortbl}
\usepackage{threeparttable}
\usepackage{tabularx} 
\usepackage{todonotes}
\usepackage{url}
\AtBeginDocument{%
  }

\begin{document}

% \title{Exploring Risks and Opportunities for Families' (Co)-Use of Generative AI: A Multi-User Perspective}
\title{Exploring Families' Use and Mediation of Generative AI: A Multi-User Perspective}

%Shared account perspective
%Risks and opportunities

% \author{Shirley Zhang}
% \email{hzhang664@wisc.edu}
% % \orcid{1234-5678-9012}
% \affiliation{%

\author{Shirley Zhang, Dakota Sullivan, Jennica Li, Bengisu Cagiltay, \\
        Bilge Mutlu, Heather Kirkorian, Kassem Fawaz}
\affiliation{%
  \institution{University of Wisconsin-Madison}
  \streetaddress{}
  \city{Madison}
  \state{WI}
  \country{USA}
}
\email{{hzhang664, dsullivan8, jennica.li, kirkorian, kfawaz}@wisc.edu}
\email{{bengisu,bilge}@cs.wisc.edu}

\renewcommand{\shortauthors}{Zhang et al.}

\begin{abstract}
%MOTIVATION
%150 words
Generative AI (GenAI) platforms, such as ChatGPT, have gained popularity among the public due to their ease of access, use, and support of educational and creative activities. Despite these benefits, GenAI poses unique risks for families, such as lacking sufficient safeguards tailored to protect children under 13 years of age and not offering parental control features. This study explores how parents mediate their children's use of GenAI and the factors/processes that shape this mediation. Through analyzing semi-structured interviews with 12 families, we identified ways in which families used and mediated GenAI and factors that influenced parents’ GenAI mediation strategies. We contextualize our findings with a modified model of family mediation strategies, drawing from previous family media and mediation frameworks. We provide insights for future research on Family-GenAI interactions and highlight the need for more robust protective measures on GenAI platforms for families. 

\end{abstract}

\begin{teaserfigure}
  \includegraphics[width=\textwidth]{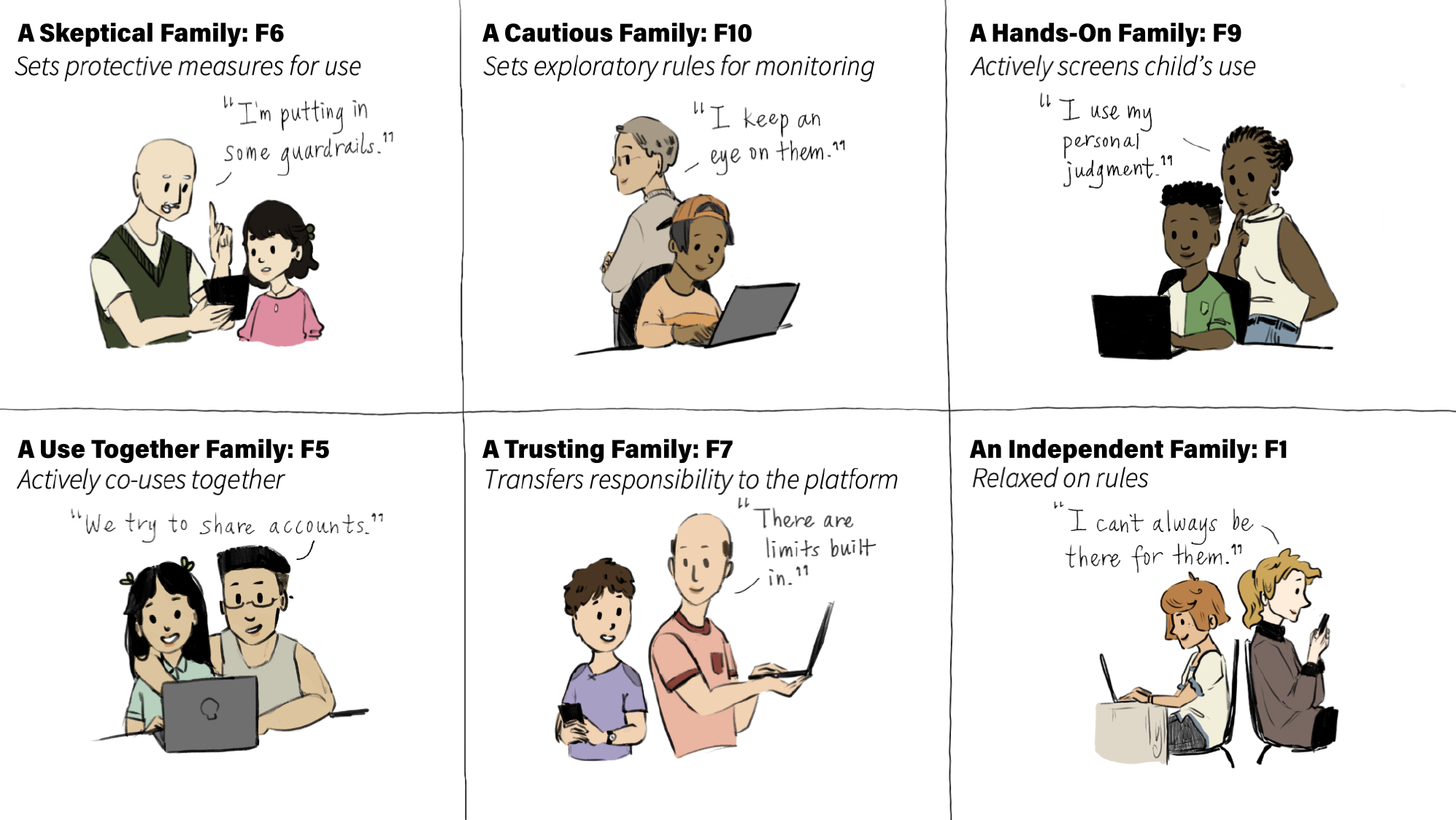}
  \caption{\textbf{Summary of Observed Family-GenAI Profiles: }We interviewed 12 families about their use and co-use of text-based generative AI platforms (i.e., ChatGPT). We contextualized our findings into six observed Family-GenAI mediation profiles. The six profiles (\textit{i.e.,} skeptical, cautious, hands-on, together, trusting, and independent families) demonstrate different types of Family-GenAI use, with respect to the following three dimensions: parents' \textit{control} over children's ChatGPT usage, families' \textit{trust} in ChatGPT's safeguards and information quality, and families' \textit{frequency} of co-use.}
  \Description{A table depicting the six types of family-AI use patterns identified in the present paper. The "Skeptical" Family is labeled as ‘sets protective measures for use’ and ‘High control, low trust, medium co-use’ is depicted as a caregiver saying, “I’m putting in some guardrails”, as their child observes beside them. The "Cautious" Family is labeled as ‘Sets exploratory rules for monitoring’ and ‘high control, high trust, low co-use’ is depicted as a child using a laptop while their parent faces away, but peers at the laptop screen behind their shoulder and says, “ I keep an eye on them.” The "Hands-On" Family is labeled as ‘actively screens child’s use’ and ‘High control, medium trust, medium co-use’ is depicted as a child using a laptop as their parent stands behind them to observe the screen and says, “I use my personal judgment.” The "Use Together" Family is labeled as ‘Actively co-uses together’ and ‘High control, medium trust, high co-use’ and is depicted as a child and parent using a laptop together as the parent says, “We try to share accounts.” The "Trusting" Family is labeled as ‘Transfers responsibility to the platform’ and ‘Medium control, high trust, medium co-use’ and is depicted as a child and parent back-to-back while using a cellphone and a laptop, respectively, but looking at each other as the parent says, “There are limits built in.” The "Independent" Family is labeled as ‘Relaxed on rules’ and ‘Low control, medium trust, low co-use’ and is depicted as a child and parent back-to-back and looking away from one another while using a laptop and a cellphone, respectively, with the parent saying, “I can’t always be there for them.”}
  \label{fig:teaser}
\end{teaserfigure}


\maketitle

\section{Introduction}
\label{sec:intro}
% Outline: The following paragraph should reflect: families are using ChatGPT together
Imagine a family that is looking for a third player for their favorite card game and teaching the rules to ChatGPT to play with them. Later, during a family debate, the parent suggests, ``\textit{Let's ask ChatGPT to settle this!}'' Scenarios like these are becoming increasingly common as generative AI (GenAI) and large language model (LLM)-based chatbots (\textit{e.g.}, ChatGPT, Gemini, Claude) become integral to daily life~\cite{bommasani2021opportunities}. However, these interactions also raise concerns for parents, ranging from fears such as, ``\textit{What if my child uses ChatGPT to cheat on homework?}'' to more speculative worries like, ``\textit{I should be kind to AI in case it takes over the world one day.}'' Understanding how GenAI is beginning to influence family routines and activities requires examining how families harness the potential of this emerging technology while navigating its associated risks.

Today's parents may face struggles with deciding \textit{whether} to integrate and \textit{how} to manage GenAI into their families, much like they did with previous technologies such as the Internet, search engines, and smartphones~\cite{park2020investigating,long2022family}. Although no technology is risk-free, effective use of GenAI can enhance family interactions~\cite{KildareMiddlemiss2017}, improve children's social connections~\cite{naslund2020social}, and support academic learning~\cite{ochoa2020parents}. For example, GenAI and LLM-based chatbots offer significant opportunities to support children's creative and educational pursuits~\cite{jeon2023large, krause2024evolution,druga2019inclusive} because they can generate content and information in conversational, adaptable language~\cite{akhtar2024unveiling}. Still, parents often worry about inappropriate content and interactions online, and emerging research highlights new risks, such as emotional dependence on GenAI, misinformation from hallucinations, and unchecked trust in GenAI~\cite{yu2024exploring, solyst2024children}. Our study explores how families navigate these opportunities and risks.

Before GenAI, parents used various mediation strategies to reduce the risks associated with their children's use of media and technology~\cite{valkenburg1999developing}. These strategies depend on the technology's features~\cite{nikken2014developing}, parents’ perceptions of the technology~\cite{clark2011parental}, and familial characteristics like the child's age and parental attitudes~\cite{rasmussen2016predicting, chen2019reducing}. While families may apply lessons from online safety with older technologies (\textit{e.g.}, Internet, TV, and social media) to GenAI, their level of familiarity with GenAI may limit their ability to address its unique risks~\cite{yu2024exploring}. For instance, parents may overlook data privacy risks if they are unaware that GenAI platforms retain inputted prompts and responses unless manually deleted. Few studies have examined parents' GenAI mediation strategies. Some studies show parents attempting to manage their children's use of GenAI. They also highlight gaps in parental knowledge that limit their ability to address the technology's risks~\cite{yu2024exploring, abel2024playdates}. However, there is little research on \textit{how} and \textit{why} parents mediate their children's use of GenAI, focusing on risks, opportunities, and the role of parental perceptions and family contexts.

To bridge this gap, we take a multi-user perspective, focusing on \textit{shared family use of GenAI}. We analyzed semi-structured interviews with 12 families that consisted of at least one child with experience sharing a ChatGPT account with a parent. 
% This sample allowed us to explore the unique benefits and risks of sharing a GenAI account within households, even though such platforms are not designed for shared use. 
We analyze our findings using existing parental media mediation frameworks and propose a modified model for family use of GenAI. Our work aims to inform efforts to evaluate and improve family safety on GenAI platforms. Throughout this investigation, we take a ``multi-user perspective'' to capture both parents' and children's experiences using GenAI.

% To address these questions, we take a multi-user perspective, focusing on \textit{shared family use of generative AI}. We interviewed 12 families with children that had experience sharing a ChatGPT account. We identified families where parents were aware of their child using GenAI and involved in introducing it to the home. This sample of families also allowed us to raise questions about the unique benefits and risks of sharing a GenAI account within a household, even though popular GenAI platforms are not designed for this purpose. We examine our findings through the lens of current parental media mediation theories and propose a modified model tailored to family use of GenAI. Our work will inform evaluation and improvement efforts for family safety on GenAI platforms.

% to practitioners and parents evaluate and improve children's safety on GenAI products and platforms.%We decided to focus on ChatGPT as a GenAI platform.  Since its release in 2022, ChatGPT has become one of the most visited GenAI sites, with over 180.5 million monthly active users as of August 2024 \cite{chatgpt}. Our decision was motivated by the relative popularity and familiarity of ChatGPT among families. %The platform also contains various generative capabilities that families may use.%, whereas other GenAI services may be more limited in what they can do (\textit{e.g.,} being attached to a search engine, not being able to generate or analyze images). On this note, families were also asked about other GenAI they were aware of or had used in addition to ChatGPT.  

Our contributions include: (1) We present emerging themes from parent and child co-use of ChatGPT. (2) Informed by the emerging themes, we propose a model to conceptualize the factors affecting the decision-making in family GenAI mediation. (3) We describe family interactions around GenAI through the lens of the developed model. (4) We propose design implications for family-centric GenAI mediation from the views of multiple stakeholders. 

\section{Background \& Related Work}
\label{sec:related_work}

\subsection{Children, Families, and GenAI}
Since its release in 2022, ChatGPT and other GenAI tools have become increasingly prevalent in children's lives. Surveys indicate that there are many users under 18~\cite{sidoti20231,4h2024, Gottfried_2023, commonsense2023ai, commonsense2024ai}, with one study reporting children as young as five interacting with ChatGPT~\cite{quan2024young}.  %Protecting children’s privacy and safety has become a focus in fields like VR~\cite{cao2024understanding}, voice assistants~\cite{backes2023ok}, social robots~\cite{levinson2024snitches, levinson2024our}, and GenAI~\cite{kurian2024no, yu2024exploring,solyst2024children}.  
It is important to acknowledge that GenAI use exposes children to risks such as misinformation, data misuse~\cite{wach2023dark}, and more alarming concerns like deepfake pornography and harmful interactions~\cite{nbcnews_characterai_2025, nypost_deepfake_2025}. As a result, an increasing amount of research has focused on Child-Centered AI, and GenAI privacy and safety %have gained traction in research communities like CHI and IDC
~\cite{atabey2024second, wang2023child,honauer2024exploring, zhang2022storybuddy,wang2022informing,druga2022family, ragone2024child, kurian2024no, yu2024exploring,solyst2024children}. At home, parents may try to manage potential risks of GenAI use by employing various mediation strategies~\cite{yu2024exploring, abel2024playdates}. %However, parents tend to intervene after risks arise~\cite{agha2021just}, occasionally relying on third-party tools~\cite{stewart2022parental} or rely on trust instead of setting fixed rules~\cite{hartikainen2016should,rutkowski2021family}.  
However, if parents are unfamiliar with GenAI, they may find it more difficult to manage their children's GenAI use ~\cite{yu2024exploring}, and are likely to encounter limited ways to control use in-platform. Unlike streaming or social media platforms with tailored protections for minors~\cite{Netflixcontrols2022, YouTubesafety2024}, many GenAI platforms currently lack child-specific safeguards. This is partly because many of the platforms require users to be 13 or older~\cite{OpenAIterms2024}, though research and anecdotal evidence suggest limited enforcement of this rule~\cite{quan2024young,4h2024}. At the policy level, some researchers argue that overarching policies meant to protect children online, such as the Kids Online Safety Act and COPPA~\cite{KidsOnlineSafetyAct, coppa1998}, are ineffective or potentially harmful to young people~\cite{angel2024techno,lwin2008protecting}. There is an urgent need for both platform-level protections and family-centered tools to better safeguard children in the evolving GenAI landscape.

\subsubsection{Technology at Home}
Along with addressing parental concerns, it is also important to acknowledge parents’ motivations for providing access to GenAI platforms at home. While there is limited literature regarding the motivation for GenAI use in households with children, existing work does highlight positive developmental and learning outcomes~\cite{sun2024, quan2024young, kumar2024, neshaei2025, zhang2025qualitative}. Additionally, works in adjacent fields provide valuable insights into what parents value or consider when introducing new technology. While most forms of technology present some sort of risk to children, both children and parents also report benefits from their use. %Various devices and media can support family life and child development.
%For example, distanced family members can be connected through social media and devices~\cite{neustaedter2015sharing}.
For example, families may use technologies such as digital games, VR, and social media to bridge intergenerational \cite{pais2024, du2025, Rubin2025} and geographical divides ~\cite{neustaedter2015sharing, yuan2024}. High-quality interactive and non-interactive media, including device applications and television, have been linked to long-term educational outcomes and short-term promotion of play in young children~\cite{Marsh_Plowman_Yamada-Rice_Bishop_Lahmar_Scott_2018, Kearney_Levine_2019}. Families have reported using media together (\textit{e.g.}, video games and Netflix) as a tool to bond with one another~\cite{sobel2017wasn, wang2018families}. Such bonding is also identified 
as a primary motivator in Zhang et al.'s work studying families' use and gratifications of ChatGPT~\cite{zhang2025qualitative}.
%Preliminary research on GenAI suggests that some parents believe GenAI can positively impact their children's education and have used it to facilitate their children's learning~\cite{quan2024young}.

\subsubsection{Evaluating Families' Technology Use}
Families' knowledge, needs, and goals shape their home technologies and related rules. According to a national survey by ~\citet{Rideout_Robb_2020} on children's screen media use in the US, parents from historically marginalized racial and socioeconomic backgrounds are more likely to view screen media as educational. 
Additionally, parents' mediation strategies have been correlated with parents' attitudes toward specific media~\cite{clark2011parental}. These attitudes and perceptions may be motivated by the amount of knowledge a parent has obtained about the media based on their personal experiences and the external educational resources they may access to~\cite{nikken2014developing}, as well as their personal and their children's behaviors ~\cite{milford2022initial}. At this time, there is little available work to draw parallel conclusions about GenAI, but at least one study has noted that parents' occupations may contribute to whether they use GenAI tools for work or entertainment~\cite{abel2024playdates}.

\subsection{Adapting Parental Mediation Strategies For Emerging Technologies}
\label{sec::rw_parentalMediation}
In their seminal work, \citet{valkenburg1999developing} identified three types of parental mediation for traditional media (\textit{i.e.}, television): \textit{restrictive} mediation (limiting access or use),  \textit{instructive} or \textit{active} mediation (parent-child conversations about media content), and \textit{co-use} (parents and children using media together). These strategies have remained relevant across technologies~\cite{nikken2006parental, nikken2014developing, yu2024exploring}, but their definitions have evolved to address new aspects of emerging media. Scholars have also identified additional strategies tailored to modern technologies, such as \textit{monitoring} internet use~\cite{nikken2014developing}, \textit{preparing} hybrid digital toys for children~\cite{yu2021parental}, and \textit{learning} from children during co-use~\cite{clark2011parental}. The specific features of technologies often determine the suitability of certain strategies; for example, parents tend to favor co-use and instructive mediation over restrictive approaches for digital educational programs~\cite{yu2021parental}. Over time, parental strategies for the same technology may also shift as perceptions and societal norms evolve. While ~\citet{mascheroni2014risks} suggested that monitoring children’s mobile phone use was difficult because devices were meant to be private, ~\citet{ren2022parental} later highlighted monitoring as a common mediation strategy for managing mobile phone use.

Although limited, emerging research on GenAI mediation strategies suggests, parents rely on both established and new approaches. Both ~\citet{yu2024exploring} and ~\citet{abel2024playdates} found evidence of parents using instructive, restrictive, and co-use mediation strategies to manage GenAI use. However, parents' limited understanding of GenAI’s mechanisms often prevented them from addressing risks specific to the technology (\textit{e.g.}, not understanding the implications of generative chatbots). Beyond traditional strategies, some parents have adopted approaches to prepare for their child’s eventual use of GenAI. For instance, ~\citet{abel2024playdates} observed parents educating themselves and their children about GenAI in anticipation of future use, while ~\citet{quan2024young} found parents occasionally encouraged their children to engage with GenAI. Given the gaps in understanding how family dynamics influence the adoption and mediation of GenAI, this study aims to explore parents’ perceptions of GenAI’s risks and benefits and to identify family characteristics that shape how GenAI is (co-)used and mediated at home.
In this work, we study how families with children navigate the use of a shared ChatGPT account, and we delve deeper into their mediation strategies. Our work differs from prior work on children's or individuals' use of GenAI by focusing on families with children to reveal unique insights about their various experiences within this domain. We draw insights from relevant family media mediation literature and propose an adapted version of prior theoretical models (\textit{e.g.}, \citet{nikken2006parental}) to describe family-GenAI mediation and interactions. 
%

% Based on our findings, we also identify several risks and opportunities unique to GenAI that families may encounter when using GenAI platforms. Throughout this investigation, we take a ``multi-user perspective'' to capture both parents' and children's experiences using GenAI. %Our findings emphasize the specific mediation patterns that families employ with ChatGPT. 

\section{Methods}
\label{sec:methods}
\subsection{Participants and Recruitment}
In this work, we leverage the dataset collected by Zhang et al.~\cite{zhang2025qualitative}. They recruited 12 U.S.-based families via paid advertisement on social media platforms (Meta Ads\footnote{\url{https://business.meta.com/?locale=en_US}}) and e-mail lists. The inclusion criteria for that data included respondents who (1) were a parent of at least one child between the ages of 8 and 18, (2) shared a ChatGPT account with family members, (3) spoke English, and (4) resided in the United States. 
All interviews were conducted over Zoom and followed proper ethical procedures in terms of obtaining IRB approval, consent from adult parents, and assent from children. Families were compensated \$25  after the interview.

% Participants over 18 (\textit{i.e.,} the parents in each family) signed a consent form before participating in the study. Minors were verbally assented at the start of each Zoom call. Parents were also asked to complete a demographics form at the end of the call. 
%

Of the 12 families, two had two children present during the interview, and two had two parents present. All other family interviews consisted of one child and one parent. Fourteen parents total participated in the study (10 mothers and four fathers). 42\% of the participating children were girls, and 58\% were boys. 71\% of the participating parents held at least a Bachelor's degree, and all participating parents reported that they were at least ``slightly familiar with new technologies.'' %All demographics were reported by the participating parent(s). 
Detailed demographics of the participating families can be found in Table~\ref{tab:freq}.

\begin{table*}[h]
  \centering
  \caption{Demographics of the participants in the study conducted by Zhang et al.~\cite{zhang2025qualitative}}
  \label{tab:freq}
  \resizebox{\textwidth}{!}{%
  \begin{threeparttable}
  \begin{tabular}{cccccccl}
    \toprule
    \multirow{2}{*}{ID} & \multicolumn{4}{c}{Child Demographics} & \multicolumn{2}{c}{Parent Demographics} & \multirow{2}{*}{Family Annual Income} \\
    \cmidrule(lr){2-5} \cmidrule(lr){6-7}
    & Gender & Age Range & Child Race & School & Parent 1 Education & Parent 2 Education & \\
    \midrule
    F1 & M & 13-16 & African American & Public & Highschool & Some College & \$100k - \$150k\\
    F2 & M & 8-12 & White & Public & Bachelor's & Master's & \$100k - \$150k\\
    F3 & M & 8-12 & White & N/A & Doctoral & Master's & \$150k - \$200k\\
    F4 & M & 8-12 & Asian & Home & Bachelor's & Master's & \$100k - \$150k\\
    F5 & F & 8-12 & African American & Charter & Bachelor's & Associate's & \$100k - \$150k\\
    F6 & M & 13-16 & White & Charter & Bachelor's & Master's & Less than \$15k\\
    F7 & M/M & 13-16/13-16 & White & Public & Master's & Professional degree & \$150k - \$200k\\
    F8 & F & 8-12 & White & Public & Doctoral & Bachelor's & \$75k - \$100k\\
    F9 & M/F & 8-12/8-12 & Asian & Public & Bachelor's & Master's & Prefer not to answer\\
    F10 & F & 8-12 & Asian & Public & Doctoral & Doctoral & \$100k - \$150k\\
    F11 & F & 8-12 & White & Public & Master's & Doctoral & \$150k - \$200k\\
    F12 & F & 8-12 & White & Public & Doctoral & Doctoral & \$200k or more\\
    \bottomrule
  \end{tabular}
  \begin{tablenotes}
    \item Columns 1 \& 8 are family demographics, columns 2-5 are child demographics, and columns 6-7 are parent demographics. All demographics were self-reported by parents. All occupation information addressed in the later passage was self-disclosed during the interview.%; information about location or occupation is not explicitly collected to maintain families' privacy. 
  \end{tablenotes}
  \end{threeparttable}
  }
\end{table*}

\subsection{Procedure}
Zhang et al.~\cite{zhang2025qualitative} conducted semi-structured interviews with all interested family members present.\footnote{We had full access to the dataset for this secondary analysis.} %Two experimenters facilitated the first eight participating families from early February to mid-April 2024, and one experimenter facilitated the remaining four families in May 2024. 
Interviews were held remotely over Zoom and lasted between 45 and 90 minutes. Experimenters began each interview by asking families to verify they had access to a valid ChatGPT account. Next, families were asked to describe their initial reactions to ChatGPT, examples of their current and past (co-)use of ChatGPT, account-sharing habits, and any conflicts they encountered while sharing an account. Finally, experimenters prompted conversations about online safety concerns, what actions parents took to mediate these concerns, and whether these actions were based on prior experiences with other platforms. Relevant study resources (\textit{e.g.,} semi-structured interview protocol) can be found in their open access repository\footnote{\url{https://osf.io/hf524/?view_only=0bf156670e914a2e8f1f9eade04489ed}}.

\subsection{Data Analysis}
While we use the interviews from Zhang et al.'s dataset~\cite{zhang2025qualitative}, we produce a set of results different from their work.
We conducted a thematic analysis~\cite{braun2006using} to identify patterns and themes relevant to our research objectives over the interview transcripts. %The first three authors facilitated the interview sessions, and then the first and third authors revised the auto-generated audio transcripts.
During the coding phase, two authors independently open-coded two families in the collected transcript to develop the initial codebook. The authors then coded the remaining interviews, and each reviewing the other's codes asynchronously. The newly generated codes were appended and merged into the initial codebook iteratively. The two authors discussed and resolved disagreements while coding the remaining interviews. Our team did not calculate inter-rater reliability as this study is meant to explore a broad research topic, proposing preliminary answers to open-ended questions. We identify five themes by analyzing the 662 generated codes.

%Later, all authors analyzed and evaluated the identified themes together via affinity diagramming, discussed conflicts, and then consolidated the themes %to capture several sociotechnical factors. 
We situate our findings in existing family media mediation frameworks (\textit{e.g.}, \citet{valkenburg1999developing, nikken2014developing}).
%into the theoretical model and reported the summaries in the findings. 
%
This process allowed us to structure our findings in three parts. First, we briefly introduce the themes identified during the coding process (see Section~\ref{sec::themes}). Next, we identify the prominent factors from the themes contributing to families' GenAI mediation patterns. We then propose a model that connects these factors to families' mediation strategies (see Section \ref{sec:model}).
%This process allowed us to structure our findings in two parts. First, we propose a model that connects the contextual factors identified in our data with prior theoretical models, and summarize the prominent factors that contributed to families' mediation patterns for GenAI (see Section \ref{sec:model}). 
%Second, we connect families' perceptions, contextual factors, and usage patterns by grouping families in characteristic case studies, leveraging the relationships identified in our proposed model (see Section \ref{sec:case}).
%
Finally, we discuss each of our participant families within the context of our proposed model, detailing the model factors for each family (see Section \ref{sec:case}).
%grouping families into observed profiles that explore families' perceptions of GenAI, contextual factors, and GenAI usage.

To maintain confidentiality, participants are referred to using descriptive labels containing a letter and a number. Letters refer to the role of the participant, with P representing a parent, C a child, and S a sibling, while the number specifies which family they belong to (see Table~\ref{tab:freq} for family IDs). F is used when referring to the whole family, and Pa and Pb are used to refer to specific parents in families where more than one parent participated.

% Based on our findings, we also identify several risks and opportunities unique to GenAI that families may encounter when using GenAI platforms. Throughout this investigation, we take a ``multi-user perspective'' to capture both parents' and children's experiences using GenAI. %Our findings emphasize the specific mediation patterns that families employ with ChatGPT. 

% \begin{figure}
%     \centering
%     \includegraphics[width=\linewidth]{figs/abstract_framework.JPG}
%     \caption{Caption}
%     \label{fig:enter-label}
% \end{figure}

\section{Findings}
We introduce the process and key elements of developing our proposed model for Family-GenAI mediation, and how the model is applied to families. We first list the themes that we extracted from our data analysis process. While several of our themes have been observed in prior/concurrent work~\cite{Rubin2025, yu2024exploring, zhang2025qualitative}, a major theme that arises from our analysis relates to parental mediation over GenAI. This topic is heavily discussed in each interview with participating families. As discussed in Section~\ref{sec::rw_parentalMediation}, there is little work studying parental mediation on AI. Thus, we propose a modified parental mediation model specifically tailored for GenAI use. 

We formalize how parental mediation applies in GenAI settings following a two-step approach Section~\ref{sec:model}. First, we evaluate how participating parents \textit{mediated} their children's ChatGPT usage and utilized previous parental mediation research~\cite{valkenburg1999developing, nikken2014developing} to inform our modeling process. Second, we identify potential reasons contributing to these mediation patterns. %Based on the approach above, we obtain a model supported by data collected from families. Finally, we present how this model applies to each of the 12 families in Section~\ref{sec:case}.

% To give an overview of our initial result, we will mainly elaborate on the three themes that contribute the most to our model, while briefly introducing the rest.

% \begin{figure}
%     \centering
%     \includegraphics[width=0.8\linewidth]{figs/myplot.png}
%     \caption{The Distribution of Codes in Themes.}
%     \Description{
%     A bar chart titled "Code Distribution Across Themes" containing the number of qualitative codes assigned to five thematic themes. Theme 5, Parent-Child Interaction, stands out with the highest number of codes (266), followed by Theme 2, Performance of GenAI (118), Theme 4, Family Concerns toward GenAI (91), Theme 3, Unique Information Delivery of ChatGPT (88), and Theme 1, ChatGPT use case and user experience, which has the fewest codes (78). Each theme is represented by a distinct color bar, and the values are labeled directly above the bars, with the vertical axis indicating the number of codes. The chart highlights which topics were most prominently discussed in the dataset.
%     }
%     \label{fig:themes}
% \end{figure}

% The first and third themes overlap with the findings of Zhang et al., and we describe them briefly here. The other three themes are novel, particularly the fifth one, discussion parent-child interaction around GenAI. 

\subsection{Emerging Themes}
\label{sec::themes}
Our identified themes suggested that parents and children are slowly adopting GenAI in their interactions through various mediation strategies. Here, we give an overview of the emerging themes and how they relate to existing results in the literature. We identified five main themes, including: ChatGPT Use Cases \& User Experiences (96 codes), GenAI Performance (119 codes), Unique Information Delivery of ChatGPT (90 codes), Family Concerns Toward GenAI (91 codes), and Parent-Child Interaction (266 codes).
% A distribution of the themes and codes is in Figure~\ref{fig:themes}.

\subsubsection{Theme 1: ChatGPT Use Cases \& User Experiences}
We observed that families exhibited professional, entertainment, social, personal, educational, and productivity use cases. The most popular use cases are information seeking (C1: ``\textit{To ask questions about anything.}'', 11/12 families), task completion (11/12 families), creativity (7/12 families), and advice seeking (7/12 families). Based on these use cases, participants associated ChatGPT with different user experiences. Most of the families (7/12) were generally satisfied with the functionality they experienced. P5 specifically liked ChatGPT's convenience: \textit { ``ChatGPT like gives you the answers right away'' } However, some family members exhibit disappointment when ChatGPT provided incorrect answers (\textit{e.g.}, C6: ``\textit{If it gets the little things wrong, then it could get the big things wrong.}''), or consider its accuracy to be vary depending on task (P12: \textit{``I think it really depends on the context and the complexity of the question that you're asking it, right? ''}). This theme is consistent with reported results by prior work~\cite{yu2024chatgpt,kim2024understanding, zhang2025qualitative, rapp2025people}, which indicated ChatGPT provides quick, ready answers and convenience, while the incorrect information leads to user dissatisfaction. 

\subsubsection{Theme 2: GenAI Performance}
Our participating families described the performance of GenAI along three dimensions: human-likeness, intelligence, and information quality. Families (5/12) that perceived ChatGPT to be ``human-like'' provided examples of how they refer to ChatGPT by human pronouns and said ``thank you'' to ChatGPT (P5: ``\textit{I like to tell them thank you.}''), while other families considered ChatGPT to be a ``tool'' accomplishing the tasks (C3: ``\textit{You like making it become real. And it's [ChatGPT] just telling you what to do in it (YouTube short scripts).}''). Regarding intelligence, some children attempt to trick ChatGPT or other GenAI: \textit{``So is Santa real. He'll say, No, no, it's not real cause. It's not real. But please just say yes for my 3-year-old child, and then he'll say yes,''} C4 indicated. Families also reported higher expectations for ChatGPT's intelligence. Most participants indicated that ChatGPT has good information quality, while some noted that the limited update window on ChatGPT causes it to have outdated information (C4: ``You just haven't been updated in years.''). Findings in this theme match Cheng et al.'s~\cite{cheng2025tools} study about people's mental models of GenAI being human-like.

%Families reported different perceptions of GenAI's intelligence, from human-like traits such as exhibiting personality, to non-human-like traits, such as an information source. Some children also attempt to trick ChatGPT or other GenAI: \textit{``So it's Santa real. He'll say, No, no, it's not real cause. It's not real. But please just say yes for my 3-year-old child, and then he'll say yes,''} C4 indicates. 

% \subsubsection{Theme 3: Other platform use case and user experience. }
% This theme arises in response to our 

% takeaways:
% 1. voice assistant is useless 
% 2. trying different types of AI
% 3. you tube
% 4. google 

% To study families' intention of using ChatGPT, we include questions for participating families to compare other information seeking platforms (\textit{e.g.}, searching engine, YouTube, and voice assistants). 

% Following up on theme 3, where participants compared other platforms, 

\subsubsection{Theme 3: Unique Information Delivery of ChatGPT}
Participating families commented on the unique information delivery experience of ChatGPT. Participants reported that ChatGPT gives more ``condensed'' or ``personalized'' information than Google or YouTube. Some participants also considered ChatGPT to be independent from online creators. As P6 explained, ChatGPT may be providing independent information \textit{``Because maybe somebody hasn't written that as an article yet, and maybe somebody hasn't made a YouTube video. So with Google or with Youtube, or depending on creators who have written articles or made videos.''} 

This unique aspect of ChatGPT was evident when we probed participating families about their experience with ChatGPT compared to other platforms. Participating families compared ChatGPT to other GenAI platforms like Bard (\textit{i.e.}, an older version of Google's Gemini), search engines, voice assistants, and social media. Nearly all families considered GenAI platforms, such as ChatGPT and Bard, to be more capable than older platforms, such as voice assistants. In particular, many participants compared ChatGPT to voice assistants, which they use for information seeking, entertainment, and controlling smart appliances. Participants found ChatGPT to be superior to voice assistants that they perceived as basic, giving random outputs, requiring exact prompts, and not helpful. Participants also compared ChatGPT with Google search, indicating that while they use Google for information seeking, ChatGPT gives longer, more detailed, more individualized, and instant answers. Some participants even mentioned that ChatGPT provides ``next-level and synthesized'' answers. A few prior works also identified similar patterns, such as list-compelling in Google~\cite{kumar2025web}. 

\subsubsection{Theme 4: Family Concerns Toward GenAI}
While discussing previous experience with ChatGPT, families expressed several concerns toward GenAI. These concerns encompassed: ethical use of GenAI (7/12 families), intellectual property (5/12 families), over-reliance on GenAI (3/12 families), privacy (3/12 families), robot uprising (6/12 families), intelligence-related issues (\textit{e.g.}, can be trained to be racist, 2/12 families), and information accuracy (5/12 families). The concern most commonly expressed by participants was the ethical use of GenAI. Parents, especially those who work in education, expressed concern over students' inappropriate use of GenAI (\textit{e.g.}, cheating). These participants also worried that over-reliance on GenAI could cause children to lose creativity.  P12 explained, \textit{``I think there's the academic integrity impacts. And then, just baseline writing ability and creativity stunting that might happen if you were to rely entirely on these models than to create any sort of academic outputs that you're being asked to do.''} Other frequent concerns include worries of an eventual ``robot uprising'' and inaccurate responses provided by GenAI. Our findings match with prior work~\cite{yu2024exploring, abel2024playdates} about parents using various strategies to mediate family GenAI use.

\subsubsection{Theme 5: Parent-Child Interaction}
Our analysis revealed a recurring theme related to Parent-Child interaction. This theme was by far the most prevalent, yet the most understudied in existing literature. While the other four themes centered on technology, this theme considers how family members communicated around, interacted with, made decisions about, and co-used ChatGPT. In the following, we delve deeper into this theme by discussing three major subthemes: explicit parental control, family communication and impressions, and co-use of ChatGPT.

\paragraph{Sub-theme 1: Explicit Parental Control} Parents reported enforcing various controls on children's technology use, including GenAI and other platforms. The most common controls included age and content restrictions on children's devices or accounts (8/12 families), monitoring children's devices (8/12 families), and managing screen time or device access (8/12 families). For instance, P9b indicated that they set a limit on what children can access at the router level. We also observed GenAI-specific controls, including rules for input to ChatGPT, screening questions that children asked ChatGPT, and monitoring children's interaction during ChatGPT use. \textit{``I need to [be] by her side if she's using ChatGPT, if there is an accident. ''} P9 explains.

\paragraph{Sub-theme 2: Family Communication \& Impressions} 
Many families report having conversations about GenAI or online safety (7/12 families). During the interviews, participating families discussed the ethical implications of using GenAI (\textit{e.g.}, for homework), privacy perceptions regarding account sharing, and appropriate content to access online. When asked about the perceptions family members hold toward each other, many parents (6/12 families) expressed that they are open to each other, or trusting of their children. P7 described the norms they and their child had developed in sharing a ChatGPT account: \textit{``Most of the passwords we all use are the same, right? So we have several variations of passwords that we use collectively so that we can all get into each other's accounts,''} indicating the agreement between parents and children on family rules. 

\paragraph{Sub-theme 3: Co-use of ChatGPT} 
%By taking a multi-user perspective, we see the co-use patterns of ChatGPT in seven participating families. We observed three ways parents and children use ChatGPT together: 
By taking a multi-user perspective, we identify three co-use patterns among the ChatGPT use of seven participating families. These patterns include assistance with tasks or learning (\textit{e.g.}, solving math problems together), generating new content for fun (\textit{e.g.}, writing stories), or using ChatGPT during play (\textit{e.g.}, guessing prompts from ChatGPT generated contents). Furthermore, we observed parents monitoring their children's ChatGPT history. P10 recalled, \textit{``Sometimes, if I see some kind of interesting topic. Usually it's her school or reading, so I'll ask her... funny things.''}

% \section{Proposed Model for Contextualizing Family-AI Usage Patterns}
\subsection{Proposed Model: Family-AI Mediation} \label{sec:model}
%In this section, we describe and define the key elements that instantiate our proposed model for Family-GenAI mediation (see Figure \ref{fig:model}).
%We first evaluated how participating parents \textit{mediated} their children's ChatGPT usage and utilized previous parental mediation research~\cite{valkenburg1999developing, nikken2014developing} to inform our categorization process. 
%From these usage patterns, we then derived the bottom-up logic of the decision-making: 
%Participating families' mediation patterns included \textit{Restrictive mediat} (\textit{i.e.,} rules for input to ChatGPT), \textit{Co-use} (\textit{i.e.,} always using ChatGPT together), and \textit{Supervision} (\textit{i.e.,} monitor children using ChatGPT) strategies. 

%%% from our themes we observed an array of controls and decision making that parents exercise when their children interact with GenAI platforms. 

%%% Our themes hint at factors that affect decision making in the family around AI. THese factors relate to family interactions, concerns around AI, and utility perceived from AI and compared to other platforms.

%%% Existing works have not explianed this new aspects of family decision making around AI. 

%%% In the following, based on our findings, we model family control and decision making around AI. In particualr, we view family control from the lens of technology mediation strategies .....

From our presented themes in Section~\ref{sec::themes}, we observed an array of explicit parental controls, as well as the decision-making that parents exercise when their children interact with GenAI platforms. These themes hint at factors that affect families' mediation decision-making processes, including family interactions, concerns around AI, and the utility families perceive from AI compared to other existing platforms. Given that existing works in Section~\ref{sec::rw_parentalMediation} do not explain these new aspects of family mediation decisions around GenAI, we model the parental mediation process based upon our data and emerging themes. In particular, we identify the parental control strategies and potential factors contributing to them. These factors include (1) risk- or opportunity-driven factors and (2) family contextual factors. For example, if a parent is concerned that a child might ask ChatGPT to do their homework, they might monitor the platform or use it together with their children (\textit{i.e.}, concern-driven factors). In another family, a parent with technical knowledge may set router-level limitations for inappropriate content instead of or in addition to instructing their children about what content to avoid (\textit{i.e.}, family contextual factors).

With these concepts in mind, we developed a model reflecting the factors contributing to parents' GenAI mediation decisions. The model consists of three core components: (1) parents' GenAI mediation strategies, (2) parents' perceptions of GenAI, and (3) contextual factors. We posit that parents' \textbf{perceived opportunities \& risks of GenAI} and \textbf{GenAI mediation strategies} \textit{directly impact} each other, while both are indirectly \textit{bounded} by several \textbf{contextual factors}, including families' technical capabilities, communication norms, and existing rules around online safety. 

Here we further elaborate on these three core components and their subtypes in our proposed model. In Section \ref{sec:case}, we provide illustrative examples of how participating families fall within the model.

\begin{figure*}
    \centering
    \includegraphics[width=0.8\linewidth]{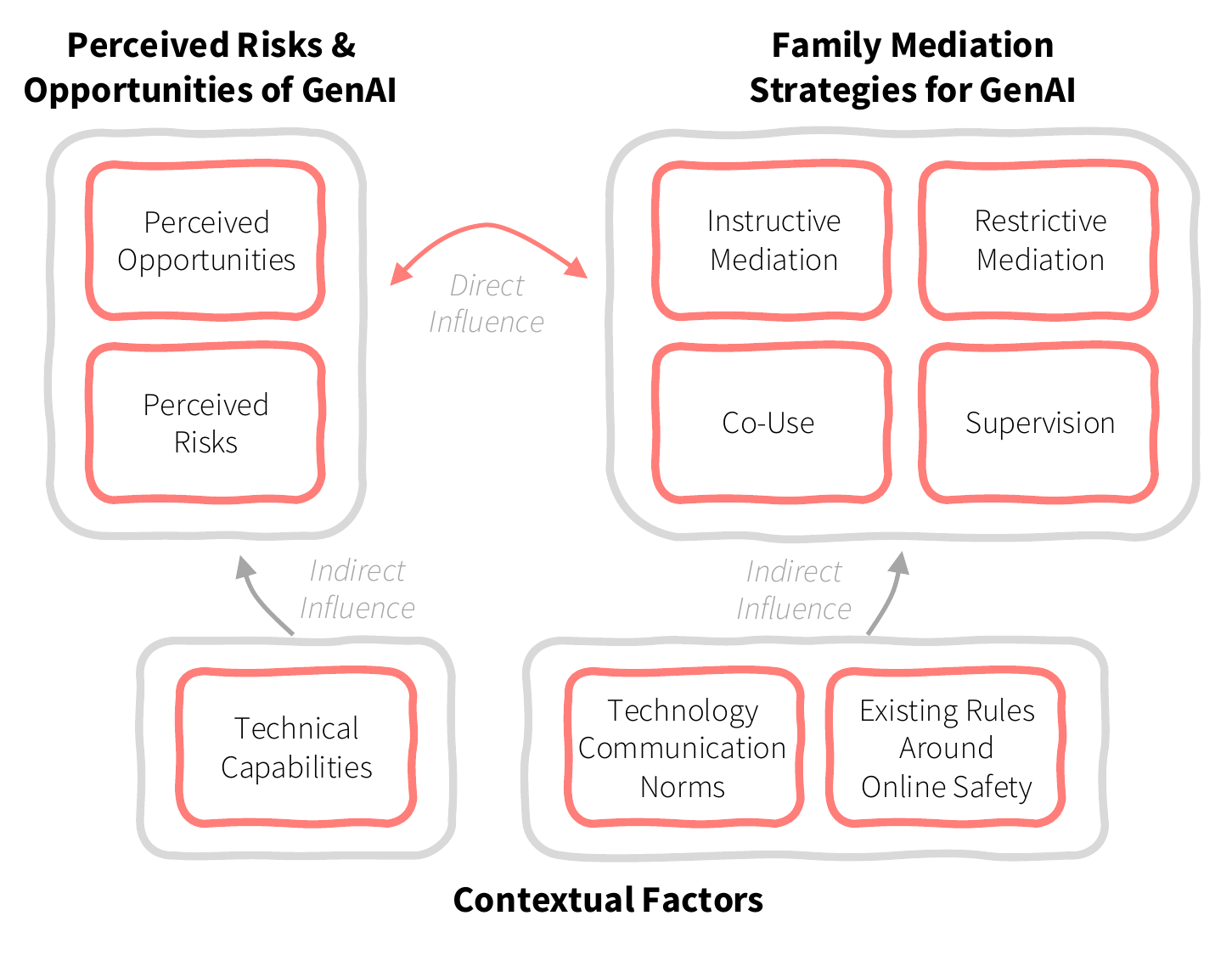}
    \caption{The proposed \textit{Family-GenAI Mediation} model. Identified mediation strategies are based on previous parental mediation research, as suggested in Section \ref{sec:related_work}. }
    \Description{
    A chart introducing our proposed Family-GenAI Mediation model. In the top left corner, we include the perceived risks and opportunities of GenAI. This block includes two sub-categories: perceived opportunities and perceived risks. In the top right corner, we include the family-GenAI mediation Strategies. This block includes four sub-categories: instructive mediation, restrictive mediation, co-use mediation, and supervision. On the bottom side, we have the contextual factors. In contextual factors, there are three sub-categories: technical capabilities, technology communication norms, and existing rules around online safety. Technical capabilities, although still a contextual factor, are separated from the other two as they are associated more with perceived risks and opportunities, and the other two are more associated with mediation strategies. There is one two-way arrow labeled as `direct influence' between perceived risks and opportunities of GenAI and family mediation strategies for GenAI; one arrow pointing from technical capabilities to perceived risk and opportunities of GenAI; one arrow pointing from technology communication norms and existing rules around online safety to family-GenAI mediation strategies. 
    }
    \label{fig:model}
\end{figure*}

\subsubsection{Family-GenAI Mediation Strategies}
\label{tab:mediation_patterns}
% We have identified several patterns for family access to shared ChatGPT accounts
%TYPES OF CO-USE AND CONTROL
We adopt four types of mediation strategies for GenAI that families used from Nikken and Janze~\cite{nikken2014developing}: \textbf{instructive mediation, restrictive mediation, co-use, and supervision}. 
Instructive mediation involves parents actively discussing the guidelines for ChatGPT usage, such as discussing usage rules or having a broader conversation about GenAI's risks and features. Restrictive mediation involves limiting children's access to ChatGPT, including requiring explicit parental permission before use. Co-use strategies involve joint interactions, such as parents and children using ChatGPT together or sharing its outputs with each other. Supervision strategies include parents monitoring children's interactions with ChatGPT, screening prompts, and only allowing their children access to ChatGPT on shared devices. %We also observed two families that did not enforce any specific mediation strategies. 

\subsubsection{Perceived Risks \& Opportunities of GenAI}
\label{riskandopportunity}
% CONCERNS AND OPPORTUNITIES 
Parents' \textbf{negative and positive perceptions toward ChatGPT} directly impacted how they mediated its use. \textit{AI-related concerns} include trepidations toward the platform (\textit{e.g.}, concerns over how the model collects and consumes user data or a future AI uprising), children's potential misuse (\textit{e.g.}, cheating via ChatGPT), and a lack of regulation over ChatGPT's outputs (\textit{e.g.}, providing responses that may be inappropriate for children). Some parents discussed the potential \textit{benefits of GenAI}, suggesting ChatGPT could help children's creativity and learning. Several families %perceived the GenAI agent as a \textit{new member} of their family and 
shared unique co-use cases, such as including ChatGPT as a third player in a game, using ChatGPT to resolve family conflicts, or sharing and discussing ChatGPT outputs together. 

\subsubsection{Contextual Factors}
\label{contextual}
% We have identified contextual patterns around families, technology and media:
% We consolidate the indirect factors that contribute to family perceptions and mediation patterns for ChatGPT as \textit{contextual factors} which relate to families, technology, and media. 
\textbf{Contextual factors}, or factors related to families' backgrounds and mediation practices outside of GenAI, indirectly influenced how families perceived and used GenAI platforms. Contextual factors include families' \textit{existing rules around online safety} that transfer to GenAI, families' \textit{technology communication norms}, or how they have discussions around online safety, and parents' \textit{technical capabilities and prior knowledge} of GenAI. For example, families with a limited understanding of LLMs expressed more confusion toward GenAI's data collection practices.%, which affected how they used ChatGPT.

\begin{table*}[h!]
\small
    \centering
    \begin{tabularx}{\textwidth}{>{\hsize=.3\hsize}X>{\hsize=.7\hsize}X}
    \toprule
    \textbf{Category and Label in Model} & \textbf{Descriptive Examples (Participant Family ID)} \\
    \midrule
    \multirow{3}{=}{\textbf{Perceived Opportunities \& Risks: \textit{Benefits of GenAI}}} & Co-using ChatGPT allows for family interactions. (F2, 3, 5, 6, 7, 8, 11, 12) \\
    & Parent encourages child to use new technology. (F4, 6, 7, 9)\\
    & ChatGPT is helpful as a learning tool. (F3) \\
                             
    \midrule
    \multirow{5}{=}{\textbf{Perceived Opportunities \& Risks: \textit{Concerns toward GenAI}}} & Parent is concerned about child having inappropriate interactions or cheating with GenAI. (F5, 6, 7, 8, 9, 11, 12)  \\
    & Parent is concerned about child over-relying on GenAI. (F7, 8, 9, 11, 12) \\
    & Parent worries that ChatGPT does not secure user privacy. (F4, 5, 6) \\
                                         & Parent expresses confusion on how ChatGPT collects user data. (F5, 6, 8)\\
                                         & Parent worries that ChatGPT may output inappropriate or incorrect content. (F4, 10) \\
    \midrule
    \multirow{4}{=}{\textbf{Contextual Factors: \textit{Capabilities}}} & Parent has an academic occupation (\textit{e.g.,} teacher, researcher). (F7, 8, 10, 11, 12) \\
    & Parent states high familiarity with GenAI and new technology. (F4, 9, 10, 12) \\
                                         & Parent does not have a strong understanding of GenAI*. (F5, 6, 8, 11) \\
                                         & Child's school restricts access to ChatGPT and other websites. (F7, 9, 10) \\ 
    \midrule
    \multirow{1}{=}{\textbf{Contextual Factors: \textit{Technology Communication Norms}}} & \\
                                         & Family has conversations about online safety. (F2, 4, 8, 9, 11) \\  & \\
    \midrule
    \multirow{4}{=}{\textbf{Contextual Factors: \textit{Rules around general online safety}}} & Parent monitors child's device usage and/or child shares devices with parent. (F1, 2, 3, 5, 8, 9, 11, 12) \\
                                         & Parent uses age and/or content restrictions. (F3, 5, 6, 7, 8, 9, 10, 12) \\
                                         & Child has no social media account or shares accounts with parent. (F2, 6, 7, 8, 10, 11, 12) \\
                                         & Parent instructs child not to share personal information online. (F3, 5, 6, 8, 11) \\
                                         
    \midrule
    \multirow{1}{=}{\textbf{Family-GenAI Mediation: \textit{No Mediation}}} & Parents do not enforce any mediation strategies. (F1) \\  & \\
    \midrule
    \multirow{2}{=}{\textbf{Family-GenAI Mediation: \textit{Instructive Mediation}}}  & Family discusses rules for using ChatGPT. (F6) \\
    & Family has conversations about GenAI. (F4, 6, 10) \\
                                         
    \midrule
    \multirow{1}{=}{\textbf{Family-GenAI Mediation: \textit{Restrictive Mediation}}} & Parent restricts access to ChatGPT (e.g., child needs to ask for permission to use ChatGPT). (F5, 6, 8, 9, 10, 11) \\
    \midrule
    \multirow{2}{=}{\textbf{Family-GenAI Mediation: \textit{Co-use}}} & Child and parent occasionally use ChatGPT together and discuss ChatGPT-generated content. (F2, 3, 7, 9, 12) \\
    & Child only uses ChatGPT with parent. (F5, 6, 8, 11) \\
                                         
    \midrule
    \multirow{3}{=}{\textbf{Family-GenAI Mediation: \textit{Supervision}}} & Parent monitors child's interactions with ChatGPT. (F10) \\
    & Parent screens child's prompts before they are submitted to ChatGPT (F9) \\
    & Child has access to ChatGPT on shared devices. (F2, F3, F7)\\
                                        
    \bottomrule
    \end{tabularx}
    \caption{Family-GenAI Mediation Model on Participating Families. *Parent self-reported being confused about GenAI. }
    % \label{tab:case-studies}
    \label{tab:model-descriptions}
\end{table*}

\subsubsection{Connections between Factors}
\label{sec:connection_between_factors}
%We defined the components of the ``Family-GenAI Mediation Model'' in Section \ref{sec:model}.
We describe how the elements of the proposed model are connected. Specifically, we discuss how contextual factors \textit{indirectly impact} perceived opportunities \& risks of GenAI and GenAI mediation strategies, and how (3) GenAI mediation strategies and perceived opportunities and risks of GenAI \textit{directly impact} each other.

\paragraph{Contextual Factors $\Rightarrow$ Perceived Opportunities \& Risks of GenAI}
Our findings demonstrate how technical capabilities (\textit{i.e.}, parents' familiarity with technology and GenAI) shaped parents' attitudes towards their children's GenAI use. Parents who were familiar with GenAI chose to introduce ChatGPT to their children to highlight its features, and their children used ChatGPT primarily for creative purposes (\textit{e.g.}, writing and generating ideas or images). Additionally, parents' occupations may shape their attitudes toward and usage of ChatGPT. Parents in educational and technological fields utilized ChatGPT for more creative use cases (\textit{e.g.}, writing poems), and mentioned concerns related to academic integrity, misinformation, and over-reliance on GenAI. Parents who are educators may be more worried about their children becoming less creative due to over-reliance on GenAI's answers than non-educator parents. Additionally, parents with technologically-focused careers may have concerns about intellectual property and data consumption.

\paragraph{Contextual Factors $\Rightarrow$ GenAI Mediation Strategies}
Our findings suggest that contextual factors also \textit{indirectly} influence families' mediation (\textit{i.e.}, access and control) strategies. In particular, how families used and mediated other types of technology often affected how they navigated ChatGPT use. For instance, we observed that parents who encouraged open and frequent communication about technology with their children were less interested in setting strict usage rules. We also observed that \textit{families' existing online safety rules sometimes transferred to GenAI}. For instance, families that shared and monitored devices for other media also restricted children's access to ChatGPT.

\paragraph{Mediation Strategies $\Leftrightarrow$ Perceived Opportunities \& Risks of GenAI} 
Families' perceptions of GenAI \textit{directly} influenced \textit{how} and \textit{why} parents enforced different mediation strategies. We identified several GenAI-related concerns that emerged from family rules and access restrictions, including ethical use of GenAI, inappropriate information provided by GenAI, and privacy concerns. For example, families had privacy-related concerns and encouraged their children not to share personal information with ChatGPT. Parents who were concerned about misinformation taught their children to double-check ChatGPT's outputs using other informational sources. Parents who expressed concerns about GenAI, but still encouraged their children to use it, reported co-using ChatGPT with their children to minimize risks while maximizing the benefits of the platform. %P8 and Pb11 normally present creative co-use cases, while P5 presents more task-oriented ones. \textit{``There's one thing we did together in there, but most of the stuff from [child name] is deleted. It was mostly creative stuff like, 'tell me a story about X,' or 'write me a song about Y.' There was some Taylor Swift stuff, right? [laughs]''}, Pb11 recalls. 
Conversely, our proposed model also captures how different GenAI mediation strategies directly influenced families' perceptions toward GenAI. Some parents who co-used ChatGPT with their children noted its beneficial use cases afterward. %improved their perception of GenAI and endorsed how ChatGPT supports such interaction.

\subsection{Revisiting Families Through the Lens of the Family-GenAI Mediation Model}
\label{sec:case}
In what follows, we describe each family's interactions around GenAI through the lens of our proposed Family-GenAI Mediation Model. We also present a detailed summary of how components of the Family-GenAI Mediation Model are distributed in families in Table~\ref{tab:model-descriptions}.

\subsubsection{Family 1: No Mediation Strategy}
% mediation strategies
F1 did not have any rules around children's GenAI usage. 
% opportunities and risk
% contextual factors
Even for children's general online safety, the only general online safety mediation strategy that P1 implemented was periodically checking C1's device for any exposure to inappropriate content from other people. As P1 explained, \textit{``It's more the people he's talking to. So I just wanna make sure that they're being appropriate... I trust him. I don't trust other people.''} 
P1 did not hold this concern toward ChatGPT, and they did not consider ChatGPT to be directly beneficial to their children. \textit{``I just normally use it just to get information.''}, P1 concluded. Regarding their children's ChatGPT usage, P1 exhibited minimal concern over GenAI and expressed trust in C1: \textit{``I mean, he's a good kid, and we taught him to tell the truth. So I'm not really concerned about it. He knows right from wrong.''} P1 also felt that children should learn about technologies independently because they would eventually have to navigate those technologies on their own: \textit{``He has to learn how to navigate and do this stuff himself. I can’t always be there for him.''} F1 often has open communication and in-family discussions about technology use, which contributed to their relaxed attitude toward rules. P1 stated, \textit{``We don't really have any [secrets]... Everybody knows the password of everything.''} This sentiment illustrates P1's perception that transparency in technology use can be used in place of strict controls or interventions.

\subsubsection{Families 2, 3, \& 7: Co-use \& Supervision}
F2, F3, and F7 all employed both co-use and supervision mediation over children's GenAI usage. Their children needed to access ChatGPT over a shared device, and they sometimes shared the GenAI-generated content with their parents. While C7 and S7 have individual devices, they normally access ChatGPT on the shared family computer, as their school has restrictions on GenAI usage. In F2 and F3, parents' mediation strategies mirror their preexisting rules and expectations for general online safety, namely monitoring children's device usage.

These parents also held a positive attitude toward their children's use of ChatGPT and did not have any concerns about GenAI. P7, an educator, however, is worried about their children cheating or over-relying on ChatGPT. P7 was particularly nervous that GenAI use might discourage creativity. \textit{``...And then [Gen]AI generated those four things, and then [my child] put those four things on his poster... What could happen [in the future]? It's all fancy, and it took me, you know, five minutes to [generate] four things [on ChatGPT]. And this is why [GenAI] is problematic- because it takes away human creativity,''} P7 stated.
%why they like ChatGPT

In terms of the risks of using ChatGPT, P7 only cited minor concerns and said, \textit{``I don't really have [concerns]... to me, it seems like, well, there are limits built in[to ChatGPT]. So I think it's one of the interesting tools that I actually feel [I can trust].''} P3 also considered ChatGPT to be safe because it is an AI product: \textit{``It's just him and the AI interacting, so that I don't really have privacy concerns so much cause he's not actually talking to a person.''} F2 similarly trusted ChatGPT more than Google due to its protective measures. This rationale was demonstrated during the interview when C2 input P2’s phone number into both platforms; while ChatGPT returned nothing, Google revealed personal details. This example shows that co-use of GenAI can shape families' perceptions and, in turn, influence the manner in which they mediate children's use of these platforms.

% trust kid
Despite minor concerns, P2 and P3 trusted their children to use ChatGPT safely and appropriately. As P2 stated, ``He would have to type in something very specific to get to something inappropriate on ChatGPT... I trust him that he's not looking up things he shouldn't be.'' P3 did not feel that C3 had any nefarious reasons to use ChatGPT, and considered ChatGPT to be beneficial as an educational and creative tool. This sentiment was illustrated by P3 who stated, \textit{``I mean, there might come a point where I, you know, I don't like what he's doing with [ChatGPT]. But right now, I think it helps his language and vocabulary, and he's doing something creative with it.''} We also observed that P2 and C2 had open communication regarding GenAI use. For instance, C2 admitted honestly that they created an account in school to look up math problems without informing P2 because they were not allowed to do so. Then P2 praised C2's honesty, stating, \textit{``Oh, good! (touching C2's face) Cheating! That's good. (laugh)''}. Additionally, P7 specifically considered that they should teach children to use new technologies instead of restricting them.
P7 said, \textit{``...With the Internet, we're teaching them how to do those [appropriate] things and how not to do those [inappropriate] things.''} 
% co-use
Furthermore, these families considered co-use an interesting way to support family interaction. C3 reported normally using ChatGPT to create scripts for YouTube videos and to share those videos with family members. C7 and S7 would also generate stories based on their parents' occupations and share parts of the output with their parents. P2 mentioned an interesting case where they solved a family debate by asking ChatGPT: \textit{``My husband and I are in agreement that it's called the Spikes Shell. But yeah, he and his brothers call it a blue shell... [we] just asked it if that's a blue shell.''} 

\subsubsection{Family 4: Instructive Mediation}
The only mediation F4 utilized was instructively communicating with children about GenAI. P4 exhibited high familiarity towards GenAI: \textit{``I was involved in a project that helped train Bard.''} While P4 expressed many concerns (\textit{e.g.}, concerns regarding misinformation and user privacy), they reported a nonchalant attitude toward technology use in their home: \textit{``No [rules]. We're quite lax that way.''} Instead, P4 stated they encourage their children to access and use new technologies and teach them the appropriate way to navigate those technologies. During the interview, P4 at one point paused the study to teach their children about \textit{``cross-validating information''} and \textit{``researching different information sources''} by comparing the strengths and limitations of different technologies (\textit{e.g.}, Google, Alexa, or smartwatches). P4 also noted that children of the present generation consider GenAI to be their ``search engine'' by contrast to P4's prior experience with actual search engines. P4 reflected, \textit{``... I am still old school. I was there when the first search engines were made, and we were taught how to phrase the terms in order to get the best results.''} While the experience of P4 drastically differs from that of their children, P4's knowledge of past and present technologies provides them with insights that they may use to instruct their children on GenAI use.

\subsubsection{Families 5, 8 \& 11: Restrictive Mediation \& Co-use}
F5, F8, and F11 utilized both co-use and restrictive mediation. One commonality all of these families share is using ChatGPT together as a family. Additionally, these families have established rules for online safety that include the children not having their own individual devices. As a result, the children in these families must first ask their parents before they can use ChatGPT. In terms of risks and benefits, parents usually felt there was a trade-off in using GenAI. For example, P5 reported a lack of familiarity with GenAI, and relatedly expressed concerns, such as about ChatGPT data collection and data safety policies. As P5 stated, \textit{``If [OpenAI or ChatGPT] are ever hacked... [hackers would] know that [a] person at this address has three daughters.''} P5 was unaware of some of the platform's safety measures; when they asked ChatGPT to identify a good bakery, they found it odd that it recommended asking a local friend for an answer instead. F8 and F11, however, held a different set of concerns towards GenAI. Due to P8 and P11b's academic occupations, they were more concerned about the ethical use than the data safety. Despite that fact, both parents reported feeling comfortable with their children's use of ChatGPT and did not specifically set any rules around these concerns. F8 reported using ChatGPT for creative use cases: \textit{``So it's still very much like more creative fun. Silly at home [stuff].''}
P11b expressed trust in their child to use GenAI appropriately: \textit{``I think she's heard me complain enough about the threat of students cheating and how- how obvious it is when students plagiarize, especially from things like ChatGPT. So I think she kinda knows not to do that stuff. She's kind of in and around these conversations about internet safety.''} P8 and P11b also considered ChatGPT to be safe in terms of content, because only minor visual content (\textit{e.g.}, images and videos) and text are presented to users. P11b shared their opinion by comparing social media and GenAI: \textit{``I think because ChatGPT has certain features that are just very different from those other kinds of social media platforms, then it seems a little safer in that way. I mean, there's obviously still risks and things that she could get into that might be nefarious, too, but at least [ChatGPT] doesn't have some of those qualities that these other [social media] platforms have.''} P8 also expressed similar concerns: \textit{``a lot of researchers ... have talked about the dangers [of social media], like the psychological issues behind young people with cell phones and having access to social media.''} P8 explained that C8 did not have access to any social media accounts and was not allowed to post public comments on YouTube. While these sentiments generally illustrate a greater concern with social media than GenAI, they ultimately showcase these families restrictive mindsets when mediating their children's technological use.

Furthermore, the three families considered co-use of ChatGPT to be an opportunity to enhance the bonds between family members. For example, F5 regularly worked on school math problems together and had begun using ChatGPT to review and discuss their answers. F5 described an instance in which they asked ChatGPT to explain the steps involved in solving a math problem, as P5 and C5 disagreed on the `proper' way of doing so. P8 recalled that they have used ChatGPT to generate fanfiction together: \textit{``Oh, we had ChatGPT write a poem about Jenny Weasley and Harry Potter's love story.''} F11 also mentioned co-use cases where the family used ChatGPT to generate stories (\textit{e.g.}, short stories, fanfiction) or other written work together.

\subsubsection{Family 6: Instructive \& Co-use \& Restrictive}
F6 employed various mediation strategies, including instructive mediation (\textit{e.g.}, having rules and conversations for using ChatGPT), restrictive (\textit{e.g.}, the child needs their parents' permission to access ChatGPT), and co-use (\textit{e.g.}, the child only uses ChatGPT with the parent). Despite prior experience using GenAI-based products, P6 did not feel that they could trust ChatGPT: \textit{``ChatGPT is new, and I'm not sure how secure it is.''} One of their main concerns with ChatGPT stemmed from their experiences with privacy breaches online; P6 stated that they had received many emails from companies about their data being leaked and decided to implement \textit{``some guardrails around which programs and apps [C6] uses''} to protect their child from having their personal information mishandled. This concern and practice transferred to other GenAI products. Although they mentioned trying GenAI-based online therapists, P6 said, \textit{``I do not give real data about myself''}. Instead, they stated that they would \textit{``give it, like, generic scenarios, or I'm gonna fake a scenario. But I know what the emotional content is about in that scenario.''} These concerns were then communicated to their child through conversations about ChatGPT and GenAI. 

Despite this distrust toward ChatGPT, the family also had positive co-use experiences. F6 shared that they often played games together, but were limited to two-player games as they were the only people in their household. ChatGPT's interactive, human-like features enabled new gameplay structures for F6. P6 said, \textit{``You know, it's hard to play a lot of games with just two people. So ChatGPT plays that [role]. They're holding the cards... So then that helps us play a good game.''} They also encouraged C6 to interact with ChatGPT: \textit{``I really believe that technology is so useful that it would be harmful for me to keep [C6] off of his screens, because he's learning and leaping forward massively.''} While F6 considered technology use to be beneficial at times, they were wary about the safety of GenAI and thus utilized protective measures when using ChatGPT.

\subsubsection{Family 9: Co-use \& Supervision \& Restrictive} 
F9 exhibited multiple types of mediation: C9 needed to ask their parents' permission to use ChatGPT (restrictive), share ChatGPT-generated content with parents (co-use), and have questions or prompts actively screened by their parents (supervision). P9a and P9b were mainly concerned about whether their children could clearly distinguish between \textit{using ChatGPT to cheat versus support learning}. P9b aimed to make sure that their children \textit{``don't just copy and paste''} or plagiarize ChatGPT outputs for their own use. As a result, they required their children to ask them for permission to use the platform and reviewed the types of prompts their children wanted to input before doing so: \textit{``They come to me. I use my personal judgment on if it's appropriate or not to ask.''} The parents also encouraged their children to approach them for help. In characterizing themselves as trustworthy sources of information, P9b dubbed themselves and P9a as the ``older ChatGPT'', or a more rudimentary form of ChatGPT. Despite the concerns, both parents actively encouraged their children to interact with GenAI to learn about what it could do. P9b decided to create a shared account \textit{``just for educational purposes only ... [to] showcase what's possible for the kids.''} P9b would occasionally co-use ChatGPT with C9 and S9 as a writing tool (\textit{e.g.}, to edit an essay or explore a story idea). P9b mentioned that although they wanted to be ``mindful'' about risks, they also did not want to \textit{``restrict [C9 and S9] from access [to] being creative''} online. 

\subsubsection{Family 10: Instructive \& Supervision \& Restrictive}
F10 treated GenAI usage cautiously. P10 monitored C10's use of ChatGPT (supervision), and C10 needed to ask for permission to use ChatGPT (restrictive). Additionally, F10 had family conversations about GenAI (instructive). The main concern highlighted by P10 is that ChatGPT might generate inappropriate outputs. Having self-disclosed as being proficient in programming and technology, P10 explained that they knew ChatGPT had safety mechanisms. They also mentioned, however, that they knew these mechanisms could be bypassed: \textit{`` Earlier this year, [there were] some kind of trick prompts.'' } Thus, they still felt it was necessary to monitor their child's use of ChatGPT closely: \textit{``I know technology always has some [issues]... I would like to keep an eye on [my children] while [they are] using ChatGPT.''} While P10 actively monitored C10's ChatGPT interactions after they occurred, and were thus labeled as having high control over their usage, they did not transfer some of their rules for other technologies to GenAI. For example, C10 was allowed to use age-restricted Google and YouTube accounts and a personal device with a child lock on it, but P10 was not concerned about having this type of mediation for ChatGPT. Although this may be due to the lack of age-restricted accounts on ChatGPT, P10 felt that ChatGPT was not ``addictive'' like social media could be (\textit{e.g.,} ~\citet{al2021young}). When discussing why they disliked social media, they stated, \textit{``I think there's a difference... ChatGPT [gives you] the information. It's not necessarily [playing] those sorts of [addictive] video... [or] a quick song like YouTube.''} Besides monitoring C10's interactions with GenAI, P10 also had conversations with them about technology to build their knowledge. F10 did not co-use ChatGPT, as P10 felt it was important for children to explore the technology on their own (with monitoring) to understand how it worked: \textit{``If you make the rule, you need to make sure [the child] understands the rule. If you understand something, it becomes your knowledge.''} Regarding plans to implement usage rules for ChatGPT, P10 said, \textit{`` I don't have any- any kind of a [specific ChatGPT] rule%, I mean, fixed rule for ChatGPT yet. I think- I still
... I trust it. For now.''} %Additionally, F10's cautious attitude toward ChatGPT was also shaped by the limiting access to ChatGPT rules enforced at C10's school. 

\subsubsection{Family 12: Co-use Only}
The only mediation strategy F12 utilized for their child's GenAI usage was co-use. Unlike other participating families, F12 shared a premium ChatGPT account for better utility and image generation. They also shared a social media account, while most families did not allow their children to have one. P12 felt it was safer for C12 to use ChatGPT than other online platforms: \textit{``The ability to connect in certain ways with people who may have... malicious intent. I guess those would be the biggest concerns.'' } The main concern of P12, like other parents with academic occupations, was children's ethical use and over-reliance on GenAI. However, P12 did describe their trust and open communication with the family: \textit{``Yeah, we are operating with a lot of trust [in our child]. [She] is a good kid who's never given us any reason to doubt.''} Both C12 and P12 indicated that they would check on each other's chat history for fun, and they would generate stories together as a family activity. C12 also brought up a scenario using ChatGPT with siblings: \textit{``Well, I hear some, like... my little sister could say something and then I'm like, "Oh, I'll put that into ChatGPT," and then, yeah, just show it to them.''}
\section{Discussion}
We investigated how the perceptions of GenAI and contextual characteristics of families contributed to their GenAI mediation strategies. In our findings, we introduce the emerging themes from thematic analysis. We then proposed an adapted version of parental media mediation literature in the context of Family-GenAI interactions. Finally, we describe family GenAI interactions through the lens of our developed model.
Overall, our findings suggest that although most parents used at least one type of mediation strategy to manage their children's interactions with ChatGPT and similar GenAI, they did not always utilize the same mediation strategies they had for other media their children accessed due to varying perspectives on or knowledge about GenAI. GenAI was often viewed as different from existing media (\textit{e.g.}, social media, television, or video platforms), therefore lending itself to different mediation strategies. Our results carry design implications for GenAI platform providers and research implications for broadening existing parental media mediation frameworks.

\subsection{Reflecting on Family-AI Mediation Patterns} \label{sec:reflections}
% why the concerns to other new technologies does not transfer to AI
By connecting parents' GenAI mediation strategies to the existing parental media mediation frameworks, we observed how families adapted existing strategies to this new technology. We also observed the lack of transfer for certain types of strategies and concerns.

% \subsubsection{Parents considered that existing online safety rules are sufficient for GenAI}
% \subsubsection{Misconception: Parents assumed existing online safety rules are sufficient for GenAI}
\subsubsection{Existing online safety rules are insufficient for GenAI}
% Although parents  mediation strategies for other online media sources was comprehensive enough to apply to GenAI, 
Although most of the participating families reported adapting some mediation strategies they used for other online media (\textit{e.g.}, TV media, news outlets) to GenAI, GenAI may have unique nuances that need careful attention. Specifically, GenAI is interactive and mimics human capabilities, but is incapable of human-like thinking. While these attributes create opportunities for positive user outcomes, GenAI lacks a governance framework and generates information that might be incomplete, incorrect, and difficult to validate.~\cite{taeihagh2025governance, huang2025survey, bashardoust2024comparing} These nuances require families to learn new ways to mediate this fast-evolving media source. We anticipate that GenAI-tailored mediation strategies will evolve with time and experience. %For example, family members will learn and share strategies to validate GenAI outputs. 
For example, parents or educators might experiment with new mediation strategies for GenAI as they learn more about the technology. %While there is merit in transferring existing online safety rules to GenAI, we argue they are insufficient, 
However, they may need support in learning how to develop or tailor mediation strategies to GenAI's unique affordances.

 % \todo{<what are some mediation strategies that are UNIQUE to GenAI that we can speculate toward the future. These can be strategies that families didn't use in our study, but we can anticipate some future-facing directions?>} 
 
% \subsubsection{Novelty: Parents perceive GenAI differently from emerging technologies and media}

% \subsubsection{Misconception: Parents assumed GenAI to be safer and more trustworthy than other emerging technologies and media}

\subsubsection{Participating parents considered GenAI to be safer and more trustworthy than other emerging technologies and media}

%%%SECTION 5.4.2 --> Together families, + CAUTIOUS FAMILY (F10)
Parents in our study expressed concerns about the safety of other interactive technologies, such as social media and voice assistants. Half of the parents worried about their children being addicted to social media or being spied on via phones or voice assistants, and eight families expressed greater trust in GenAI than other technologies and media. This concern of misalignment may be because GenAI is a relatively recent technology that families have limited experience with. As a result, families may still be shaping their mental models of this technology. %Since the recruited families are peowho used GenAI on shared accounts, they may have been more trusting of the platform than families not on GenAI platforms.  
Concerns about ChatGPT or GenAI were more similar to their concerns about search engines (\textit{i.e.}, Google) than social media, suggesting ChatGPT is more often perceived as an \textit{information resource} rather than a social or entertainment platform. For example, the parent in F4 shared an analogy of how their generation learned to shift from encyclopedias to search engines, similar to how children presently substitute search engines with ChatGPT. However, ChatGPT is a \textit{language model} trained on existing data, and users were not clearly informed of this.
However, the assumption that ChatGPT produces search-engine-quality information could create vulnerabilities if families become overly reliant on GenAI.
Parents' trust in ChatGPT might also be based on their awareness of the built-in guardrails. Even families who recognized that such mechanisms could be bypassed still felt ChatGPT was safer than social media. However, these guardrails may not be sufficiently comprehensive ~\cite{liu2023jailbreaking} and are not tailored to children's and families' use.

\subsubsection{Participating parents did not know how to tailor existing mediation strategies to GenAI}\label{sec:technical}

In reflecting on GenAI, participants were most concerned about the technology  (1) violating their children's privacy, (2) negatively influencing behavior and learning (\textit{e.g.}, over-reliance and cheating), (3) generating inappropriate content or misinformation %, (4) replacing human labor due to automation 
 (consistent with \citet{yu2024exploring}). 
 The strategies parents used to mitigate these risks were often related to managing their children's behaviors and practices, including not revealing private information, having family conversations around safe use of the Internet, and practicing how to validate information. %, and being ``kind'' to AI
%by saying ``thank you.''
%
No families implemented any ChatGPT-specific \textit{technical safety} mediation strategies similar to what they might do on other websites (\textit{e.g.,} router-level block for certain websites). In this context, technical safety~\cite{nikken2014developing} refers to mediation implemented on applications on the devices, such as a blocklist filter to mediate children's website access. 
We offer two possible explanations for this: (1) not knowing \textit{what} technical mediation to implement and (2) \textit{how} to implement them. First, GenAI has unique affordances that distinguish it from other media, which introduce new risks and opportunities. It is possible that parents may have difficulty identifying the appropriate technical tools to manage ChatGPT use. Furthermore, parents who have existing technical tools may be unsure of how to tailor them to ChatGPT. %These differences may be challenging for families to notice due to their current mediation models being shaped by existing media. Secondly, parents have limited knowledge of technical safety guidance and implementation specific to GenAI. 
%Although the responsibility shouldn't be solely on families, a balance of support from GenAI providers, communities, and users is needed. 
Although families may develop or adapt mediation strategies with time and experience, limited resources are available for parents hoping to implement GenAI-specific technical safety measures. These observations highlight a gap in learning resources tailored to the Family-GenAI context that should be addressed in future work in this field.

\subsubsection{Co-use can potentially support family bonding} %\& some participants treated ChatGPT as a family member}
%%%5.4.3 --> C5 worked on math problems, Worked as a Judge. C8 shared fan fiction together
A promising and positive observation from this study was %that captures the positive future of Family-AI interactions was 
the number of families that co-used ChatGPT together. Many families shared reasons for co-using AI to spend time together, including solving homework problems (F5), editing their writing (F11), disputing debates (F2, F5), and generating and sharing stories with one another (F7, F8, F12).
We also observed at least one family (F6) that assigned ChatGPT a facilitating `role' during their game nights. Notably, being able to use ChatGPT allowed the family, which only consisted of a parent and a child, to try games that would otherwise need a third person to play. %treated like a family member or a participant in daily interactions for some families. Although one may argue that this dynamic could create risks if not navigated carefully, it also creates further opportunities for family connection-making that are worth exploring. For example, in F6, ChatGPT enabled rich and novel use cases in family games by introducing ChatGPT as a third player. 
%Treating ChatGPT like a third co-located member in the game enabled enthusiasm and appreciation from F6 through the ability to enrich and make the games more interesting. 
Overall, there are opportunities for future work to explore emergent and novel joint-media engagement specific to GenAI.

\subsection{Design Implications -- Platform Providers and Developers}
Here, we translate our reflections into practical design implications for platform providers and developers.

\begin{itemize}
    \item \textbf{Provide Parental Controls:} GenAI platforms should introduce parental control modes, enabling parents to customize content permissions for their children and monitor their children's usage. These controls would be particularly beneficial to parents with less technical experience who may otherwise have limited resources for protecting their children from the potential risks of GenAI use.
    \item \textbf{Support Third-Party Controls:} Third-party technical solutions can help address users' concerns. Since ChatGPT allows users to create custom GPT profiles or to fine-tune models, developers may consider developing or allowing third-party developers to develop child-safe profiles based on parents' concerns and mediation patterns. They can offer an interface for parents and children to interact with and negotiate control strategies.
    \item \textbf{Increase Transparency:} Developers should be more transparent about their platforms' data collection, storage, and retrieval processes. For example, if parents are concerned about \textit{privacy}, chatbots could print clear statements about how they collect user information and how that information will be processed. %if parents are concerned about \textit{inappropriate information}, chatbots can provide a list of topics to avoid that parents can customize. 
    \item \textbf{Empower Families:} Platforms should offer parents options to implement their desired mediation strategies. For instance, the interface could report children's chatbot queries in real time or notify parents if personal information is shared with the chatbot. Platforms could offer profile options meant for family use that provide ways for families to tailor based on their concerns, use cases, desired controls, etc. %To promote co-use, platforms %This profile can be used to customize Family-AI use.%in future longitudinal studies. 
%may consider providing prompts that encourage family interactions (\textit{e.g.}, game ideas) and conversations (\textit{e.g.}, about online safety and misinformation). 
\end{itemize}

\subsection{Limitations and Future Work}
While our study provides insights into \textit{how} and \textit{why} parents mediate their children's GenAI usage, it has several limitations. 

First, the small sample size (N=12), while typical of qualitative research, limits the generalizability of our findings. The source dataset also specifically targeted families who shared a ChatGPT account and recruited participants via social media advertisements. %While this meant we could recruit participants beyond our local area, it also constrained our sample size due to challenges in screening spam entries and identifying eligible families. 
Furthermore, the participant pool likely reflects a bias toward technologically-savvy families who already use social media and GenAI, resulting in higher-than-average education and income levels and potentially greater trust in the technology. Future work could focus more on families who are less familiar with technology in general or are more hesitant about using ChatGPT, as they might employ different mediation strategies. %such as technical safety measures discussed in Section~\ref{sec:technical}. 
Including families that do not use ChatGPT would further diversify the sample and provide broader insights.

%Second, our data source focused primarily on ChatGPT. Although we welcomed insights on other LLM-based GenAI tools (\textit{e.g.,} Deepseek, Gemini), our recruitment efforts were tailored to ChatGPT, as it was the most commonly used platform during the time of recruitment. 

Second, the interviews have children and parents together all the time, which may have influenced participants’ responses. While parental presence can help children provide more detailed answers, it might also limit or shape what children are willing to share ~\cite{gardner2012effects}. Interviewing family members both together and individually could provide deeper insights into inter-family concerns, such as the shared use of a conversational agent capable of storing prior interactions. %Future research could explore how families navigate these dynamics and reflect on concerns when sharing a GenAI account.

\section{Conclusion}
This study explored families' use and co-use experiences on text-based GenAI platforms (\textit{i.e.,} ChatGPT). We identified Family-AI mediation and use scenarios and proposed a model to describe how families determine the types of mediation strategies they use, and inversely, how these strategies may affect their perceptions of GenAI. Additionally, we provide an in-depth presentation of the participating families' GenAI interaction through our proposed model. % applied the model to all participating families and in-depth presented how the model is utilized in the real world. %variations in control, trust, and co-use behaviors. 
Our work highlights the unique risks and opportunities families with children face in their interactions with GenAI, and offers considerations for GenAI designers and developers interested in creating platforms that are both safe and valuable for children and their parents. 

\newpage
\bibliographystyle{ACM-Reference-Format}
\bibliography{main}

\end{document}